# Alain Aspect's experiments on Bell's theorem: A turning point in the history of the research on the foundations of quantum mechanics[#]

Olival Freire Junior

Instituto de Física – Universidade Federal da Bahia

Brazil

E-mail: freirejr@ufba.br

Abstract: Alain Aspect's three experiments on Bell's theorem, published in the early 1980s, were a turning point in the history of the research on the foundations of quantum mechanics not only because they corroborated entanglement as the distinctive quantum signature but also because these experiments brought wider recognition to this field of research and Aspect himself. These experiments may be considered the most direct precursors of the research on quantum information, which would blossom a decade later.

## 1. Introduction

During some time of the 20th century, the research on the foundations of quantum mechanics was poorly considered among physicists. This was true, particularly concerning the possibility of changing quantum mechanics with additional variables. Evidence of this *Zeitgeist* abound. David Bohm's alternative interpretation of quantum mechanics was considered to belong "to the philosophy of science, rather than to the domain of physical science proper" (Messiah 1961, p. 48). Hugh Everett's interpretation faced such harsh opposition that he abandoned physics (Freire Junior, 2015, pp. 107-115, 129). On December 2, 1966, when John Bell had just published his theorem contrasting quantum mechanics and local hidden variables, Léon Rosenfeld wrote to him: "I need not tell you that I regard your hunting hidden parameters as a waste of your talent; I don't know, either, whether you should be glad or sorry for that." Disregard for Bell's theorem continued even when experiments to test it were being performed. John Clauser, who conducted the first experiment contrasting local hidden variables prediction with quantum mechanics, faced obstacles to obtaining a permanent position because American physicists doubted if what Clauser was doing was "real physics" (Freire Junior, 2015, p. 271).

---

[#] Paper to be published in the *European Physical Journal D* topical issue titled Quantum Optics of Light and Matter: Honouring Alain Aspect. According to the *EPJ D*, this is "a Festschrift celebrating the research career of Alain Aspect on the occasion of his 75th birthday, and coinciding with the presentation of the 2022 Nobel Prize for Physics. Edited by David Clément, Philippe Grangier and Joseph H. Thywissen, the issue comprises a set of historical and personal perspectives on Prof Aspect's career, together with a set of scientific articles on contemporary research in quantum and atom optics, providing fascinating perspectives for quantum information science and quantum technologies."

Rosenfeld's stand reflected the shared wisdom at the time that there was no physics to be done about the possibility of completing quantum mechanics with additional variables. This wisdom was grounded on the existence of a mathematical proof formulated by John von Neumann to prove that any hidden variables would be incompatible with quantum mechanics. It was also rooted in the widely shared view that Niels Bohr had put an end to this issue in the debate with Albert Einstein on the EPR experiment and the completeness of this physical theory. In fact, to counter this wisdom, Bell had criticized von Neumann's proof and maintained that the last word on the matter of the completeness had yet to be said.[1] However, while well considered for his work on high energy and accelerators, Bell was not yet the authoritative voice on the foundations of quantum mechanics that he would later become. In fact, traces of this previously shared wisdom survived the initial reception of Bell's theorem. Thus, Abraham Pais, in his biography of Einstein, assessed the EPR paper had no bearing on physics and did not cite Bell's theorem as a development of this issue (Pais, 1982, Chapter 25c).

A half-century later, the scene had completely changed. In 2010, the Wolf Prize in physics was awarded to John Clauser, Alain Aspect, and Anton Zeilinger "for their fundamental conceptual and experimental contributions to the foundations of quantum physics, specifically an increasingly sophisticated series of tests of Bell's inequalities or extensions thereof using entangled quantum states."[2] A little later, in 2013, Alain Aspect went to Copenhagen to receive the Bohr Medal, awarded by UNESCO for his contribution to understanding the non-locality of quantum mechanics, that is, entanglement. His talk at the ceremony was an exposition on entanglement.[3] Aspect explained that it is both a physical phenomenon that challenges our classical intuition and the physical ground for the current research on quantum information and cryptography. These prizes expressed both the high status acquired by work on entanglement among physicists and the role played by Aspect's work in this story.

---

[1] Bell exemplified the "view that the possibility of hidden variables has little interest," citing works by Rosenfeld, Pauli, Heisenberg, and N. R. Hanson (Bell 1966, p. 451).

Aspect's seminal experiments, the results of which were published in 1981-1982, and his doctoral thesis presented in Paris in 1983, lie midway in this story. They were major breakthroughs in our understanding of the conflict between quantum mechanics and local realism and the establishment of entanglement as part of the physics conceptual and practical toolkit. Furthermore, they were also a turning point in the recognition among physicists of how good the physics being done on the foundations of quantum mechanics was. These experiments were responsible for the early prestige Aspect was given as an outstanding experimental physicist. Finally, in hindsight, Aspect's work on Bell's theorem is a milestone in the road leading to more sophisticated experiments on entanglement and to the blossoming of research on quantum information.

The second section of this paper presents Bell's theorem and its first experiments, dealing mainly with those which used pairs of optical photons before Aspect announced his plan to perform a new investigation. Aspect's works are analyzed in the third section. The fourth section deals with the impact of his work on the physics community. The epilogue is a summary of the theoretical and experimental developments concerning Bell's theorem and the

---

[2] Wolf Fund Prize Announcement 2010, online: http://www.wolffund.org.il/index.php?dir=site&page=winners&cs=283&language=eng
[3] The conference full title was "An Open World - Science, Technology and Society in the Light of Niels Bohr's Thoughts." See http://bohr-conference2013.ku.dk/

physical phenomenon of entanglement as they have evolved since Aspect's early works.[4]

## 2. The early history of Bell's theorem

Since the very inception of quantum mechanics, around 1925-1927, when probabilistic descriptions of quantum states were introduced by Max Born, some physicists longed for a deeper theory able to overcome this weird feature. They appealed to an analogy with classical mechanics and statistical mechanics. Thus, if one considers quantum mechanics analogous to statistical mechanics, one should look to find the quantum counterpart to classical mechanics, which would demand more variables than those already used in quantum mechanics. Indeed, to counter this appeal was von Neumann's motivation with his proof against the possibility of additional variables compatible with quantum mechanics. The situation became more acute in 1935 when Albert Einstein challenged the completeness of quantum mechanics through the EPR thought experiment. However, Bohr had reacted to the challenge by showing that no inconsistency appears if one considers the physical phenomenon as the wholeness of the system under investigation and the measurement devices required for such an investigation.

In the early 1950s, David Bohm challenged both von Neumann's mathematical proof as well as its usual interpretation maintained by Bohr, Pauli, Heisenberg, and most physicists. Bohm materialized his challenge with a model of particles with well-defined paths guided by a new potential, the quantum potential, which was able to reproduce results from non-relativistic quantum mechanics. In hindsight, Bohm's feat can be explained by the fact that its quantum potential introduced non-local features, thus making it compatible with standard quantum mechanics, but this was not understood this way at the time. In addition, while Bohm was aware his model was a counterexample to von

---

[4] In the sections 3, 4 and the epilogue, I draw from my book (Freire Junior, 2015) "The Quantum Dissidents - Rebuilding the Foundations of Quantum Mechanics (1950–1990)", pp. 274-279 and 290-301.

Neumann's proof (Bohm 1952, p. 187), he did not identify flaws in this proof.[5] Thus, as late as the early 1960s, the foundations of quantum mechanics, as far as the possibility of completing this physical theory through additional variables was concerned, were in an untidy state. Nobody could say what the problem was with the coexistence of von Neumann's proof and Bohm's counterexample. In this context, EPR did not gain much attention among physicists. They thought Bohr's answer was right even though they did not analyze in detail what was at stake with such an experiment.[6] The reception of the EPR paper led Sidney Redner (2005) to label it a "sleeping beauty." It laid dormant without getting much attention and no citations until a particular day when the paper suddenly began to receive attention from the science community.

John Bell's mid-1960s papers were the equivalent of the prince's kiss in the fairy tale in the history of physics. According to his recollections, Bell became fascinated by the subject of physical models challenging von Neumann's proof in the early 1950s. "In 1952 I saw the impossible done," and "Bohm's 1952 papers on quantum mechanics were for me a revelation" were Bell's (1982, 1987) statements. However, his attention was diverted to other subjects, high energy physics, and accelerators, and only in the early 1960s he came back to von Neumann's proof. Bell's works focused on the critical analysis of the assumptions behind von Neumann's proofs and their reformulations, and later, on the assumptions behind the Einstein-Podolsky-Rosen *Gedankenexperiment*. After understanding the reasons why Bohm's model had survived such proof, it was as non-local as quantum theory; Bell asked himself what kind of conflict may remain between hidden variables and quantum theory.

Then Bell went back to the EPR experiment and took the next logical step: to isolate what reasonable assumptions were behind Einstein's argument and check the compatibility between these assumptions and quantum mechanics.

---

[5] Bell would criticize Bohm's "lack of clarity, or else accuracy" on this point (Bell 1966, p. 451).

[6] Noting this attitude, John Bell and Michael Nauenberg remarked: "We emphasize not only that our view [that quantum mechanics is, at the best, incomplete] is that of a minority but also that current interest in such questions is small. The typical physicist feels that they have long been answered, and that he will fully understand just how if ever he can spare 20 minutes to think about it (Bell and Nauenberg 1966).

Einstein's realism implied physical systems have well-defined properties independent of being observed. In addition, Bell (1964) noted that the "vital assumption" when dealing with a two-particle system is that what is being measured on one of them does not affect the other. He recalled Einstein's dictum, according to which, "on one supposition we should, in my opinion, absolutely hold fast: the real factual situation of the system S2 is independent of what is done with the system S1, which is spatially separated from the former." As Bell knew that Bohm's hidden-variable theory did not satisfy this dictum, he built a simple model of a hidden variable theory obeying Einstein's assumptions and showed that its results conflict with quantum mechanical predictions in very special cases. This is Bell's theorem: no local hidden-variable theory can recover all quantum mechanical predictions. This resulted from the violation, by quantum mechanics predictions, of inequality based on such hidden variables. Since then, many other analogous inequalities have been obtained, adopting somewhat different premises. As a result, it is usual nowadays to speak of Bell's inequalities as the quantitative measurement of Bell's theorem.

*The first experiments*

Meaningful reactions to Bell's theorem were not immediate. At the end of the 1960s, however, a few American physicists acknowledged the novelty implied by this theorem. After working on different paths, they joined efforts and produced the CHSH paper (Clauser et al. 1969). In this paper, John Clauser, Abner Shimony, Michael Horne, who was a doctoral student under Shimony at Boston University, and Richard Holt, a doctoral student of Francis Pipkin at Harvard, translated Bell's theorem into viable experiments and noted that no available experimental results could be used either to corroborate quantum mechanics or to support hidden variables based on the local realistic assumptions.

In the early 1970s, the first results of tests of Bell's theorem using pairs of polarized photons were available, and their results conflicted with each other. The conflict grabbed the attention of physicists not initially involved with the subject, such as John Archibald Wheeler, and led to a rush for new experiments.[7] At Berkeley, Clauser and Stuart Freedman obtained results confirming quantum predictions and violating Bell's inequalities (Freedman & Clauser 1972), and, at Harvard, Holt got results against quantum mechanics (Holt 1973). Clauser tried repeating Holt and Pipkin's experiment, and unlike the Harvard physicists, he obtained results confirming quantum predictions. Edward Fry and Randall Thompson attempted to improve the experimental techniques using a tunable laser to excite the atomic sample. With this improvement, Fry was able to get results in 80 minutes, while Clauser's first experiment lasted around 200 hours in data collection. Fry's and Clauser's results confirmed quantum mechanics predictions (Clauser 1976; Fry & Thompson 1976). However, so far, no experiment was immune to the loophole of a hypothetically unknown interaction between the two polarizers, which were far away from each other. Such a loophole has since been known as the locality loophole. The year before, Alain Aspect had entered the game suggesting a different experiment to test Bell's theorem (Aspect 1975), inspired by an earlier suggestion of John Bell to close that loophole, which Aspect (1976) detailed one year later.

## 3. Aspect's experiments and doctoral dissertation

Aspect's (1975) main proposal was to use "versatile" polarizers "whose orientations are changed rapidly and repeatedly in a stochastic manner" in such a way that the principle of separability holds: "the response of a polarizer is independent of the state of another device that is separated from the former by a space-like interval, according to special relativity." Thus, the time required to change the polarization orientation should be less than the time light requires to cross the distance between the two polarizers. It was not a new idea. Indeed, Aspect was following a proposal from Bell, who had suggested it at the end of the paper presenting his theorem (Bell 1964), recalling that Bohm and Yakir

---

[7] Optical photons were not the sole choice for Bell's theorem experiments. For a wider review, see Clauser & Shimony (1978). For a balance of these experiments, see Paty (1977). On Wu & Sakhnov experiment, see Silva (2022).

Aharonov had formulated such a suggestion in 1957. Clauser had thought about this experiment but did not try to perform it.

At the time Aspect made such a proposal, he had returned from the civil service (French *coopérant*) in Africa and was looking for a subject in optics for his French *doctorat d'état*. It was at the time when France had two types of doctoral degrees, the first *doctorat de troisième cycle,* and the second *doctorat d'état*. Aspect already had the former working with holography. The latter was a significant research project, longer than the current doctoral degrees, and without the need for supervision. These two kinds of degrees would be unified in 1984 into the current French Ph.D.

In 1974 Aspect had gone to the Institut d'Optique at Orsay looking for a subject for his research and met Christian Imbert there, who handed him a bunch of Bell's papers. Aspect became fascinated by them and thought about performing the experiment Bell had suggested: to rotate the polarizers while the photons were in flight. It was a daring but wise plan. Had Aspect chosen to repeat the kind of experiments being run in the U.S. the following year his proposal would have become totally unattractive. Indeed, as we have just seen, in 1976, Clauser's and Fry's results settled the initial tie between Freedman & Clauser's and Holt's conflicting results. Aspect's proposal was more ambitious but harder to be won due to the intrinsic technical difficulties of changing the polarizers while the photons were in flight. Aspect's strategy was to use acoustic standing waves to produce interactions between them and a beam of light, thus obtaining two channels with transmitted and diffracted beams.

To plan and perform experiments, Aspect spent around five years. In the meantime, he realized that the very techniques he would use could produce better results for the kind of experiments that had already been performed by S. Freedman, J. Clauser, R. Holt, E. Fry, and R. Thompson in the U.S. This led him to plan the realization of three different experiments.

Eventually, the experimental results were published between 1981 and 1982. The first experiment was a replication of the experimental test of Bell's inequalities early conducted by Clauser, Holt, and Fry. However, Aspect used two tunable lasers to excite the sample, which provided him a source of higher efficiency. The experimental running lasted 100 s. Furthermore, Aspect took the opportunity to check Furry's conjecture, which suggested that the quantum non-locality would vanish after the photons traveled the order of the coherence length of their associated wave packets, which meant 1.5 m in this experiment. In mathematical terms, Furry's conjecture says that a pure state would evolve towards a mixture of factorized states after such a distance. In the Aspect's experiment, the source was separated from the polarizer by 6.5 m. This first experiment was conducted with the collaboration of Aspect's undergraduate student, Philippe Grangier, and the research engineer Gérard Roger. In the second experiment, still with Grangier and Roger as collaborators, Aspect used two-channel polarizers to have a straightforward transposition of the EPR *Gedankenexperiment*. In previous experiments held in the U.S., when one of the detectors was not triggered, it was impossible to know whether this resulted from the low efficiency of the detectors or whether the polarizer had blocked the photon, which would be a real measurement. For this reason, auxiliary experiments without the polarizers were needed to circumvent the intrinsic deficiency of the setup.

The third experiment had the widest impact. Now with Jean Dalibard and Gérard Roger as co-workers, Aspect produced the first-ever test of Bell's inequalities with time-varying analyzers. Dalibard was then a young student at École Normale Supérieure who volunteered to work with Aspect on this experiment because he wanted "to turn the knobs of an experiment that will remain in the books," a premonitory view.[8] Aspect's ingenuity was to use a switch to redirect the incident photons to two different polarizers. This device worked through the planned acousto-optical interaction with an ultrasonic standing wave in water. Aspect was aware such a device would operate in a periodic manner and not in a genuinely stochastic way. They looked to attenuate such limitations once they could not be completely overcome.

In all these experiments, Aspect obtained results in clear-cut violations of Bell's inequalities and strong confirmation of quantum mechanics predictions. The first experimental result was $\delta_{exp} = 5.72 \times 10^{-2} \pm 0.2 \times 10^{-2}$, while the concerned Bell's inequality was $\delta \leq 0$ and the quantum mechanical prediction was $\delta_{QM} = 5.8 \times 10^{-2} + 0.2 \times 10^{-2}$. Thus Bell's inequality was violated by more than 13 standard deviations (Aspect et al., 1981). The second experimental result was $S_{exp} = 2.697 \pm 0.015$. In this case, the Bell's inequality at stake was $-2 \leq S \leq 2$ and $S_{QM} = 2.70 \pm 0.05$, which was to that date the strongest violation of Bell's inequalities ever reported. In each of these experiments, the runs lasted 100s (Aspect et al., 1982b).

---

[8] Alain Aspect, interviewed by O. Freire & I. Silva, 16 Dec 2010 & 19 Jan 2011, American Institute of Physics, College Park, MD.

The last experimental result, from the experiment using time-varying analyzers, was telling precisely because of its novelty, and possibly because of this it was the result that most resonated in the physics community despite its accuracy being less than the other experiments. For this case, the Bell's inequality being checked was $S \leq 0$ and the value predicted by quantum mechanics was $S_{QM} = 0.112$. The runs lasted 200 min and the experimental result was $S_{exp} = 0.101 \pm 0.020$, violating Bell's inequality by five standard deviations (Aspect et al. 1982b).

Jumping ahead of time, let us say that the three experiments were included in Aspect's doctoral dissertation, which was assessed by the panel of examiners at the Université de Paris – Sud at Orsay on February 1, 1983. It is larger than the ensemble of results and papers already published. It includes a clear introduction to Bell's theorem and its first experiments, an invaluable source if used as a textbook on this subject. In addition, each technical, experimental, or conceptual choice, either mentioned or briefly justified in the papers, are explicitly considered in detail here. The dissertation's conclusion deserves comments. Aspect's main conclusion is that Bell's inequalities are violated, and his experimental results are in excellent agreement with quantum mechanics predictions. However, he also presents their main limitations, namely the weak sensitivity of the photodetectors, which led to the additional hypothesis of taking the number of pairs of detected photons as a representative sample of the emitted photons, and the not strictly random nature of the changing polarizers, in the case of the third experiment. However, Aspect is not optimistic about possible improvements to such features soon, a topic we come back to in the following section.

Aspect's conclusion is cautious but strong. From a strictly logical point of view, the dispute between quantum mechanics and local hidden variables is not yet closed. However, he puts the emphasis of his conclusion on another side of the issue. The experiments are in precise agreement with quantum mechanics predictions for these factual experiments. On this issue, he ends his dissertation recalling what Bell had said about Aspect's suggested experiment. Indeed, in 1976, when Aspect announced the experiment with the changing polarizers, Bell (1976, p. 442) declared: "It is therefore of the highest interest that an atomic cascade experiment is now under way, presented here by Aspect, in which the polarization analyzers are in effect **re-set while the photons are in flight**." Aspect (1983, p. 346) further recalled what Bell had said at that time, "if this experiment gives the expected result, this will be the confirmation of what is, from my point view, considering the discussions on locality, one of the most telling predictions of quantum theory."

The panel evaluating Aspect's doctoral dissertation reflects the professional network he was able to build around him and the subject of Bell's theorem. These include André Maréchal and Christian Imbert, who welcomed him at the Institut d'Optique at Orsay for a French *doctorat d'état*. Maréchal, the director of the institute, a major figure in French optics, and Imbert, who handed him a collection of Bell's papers. John Bell and Bernard d'Espagnat were experts in the foundations of quantum mechanics who discussed and supported him from the start of his doctoral research. Franck Laloë, who had co-authored with C. Cohen-Tannoudji and B. Diu an influential quantum physics textbook (Cohen-Tannoudji et al. 1973), shared an early interest in Bell's theorem, having attended the 1976 Erice conference organized by Bell, which was one of the first conferences dedicated to this subject.[9] More recently, Laloë (2012) published an authoritative book on the subject, provocatively titled "Do we really understand quantum mechanics?" Finally, Claude Cohen-Tannoudji, a leader in French quantum optics who would share the 1997 Physics Nobel Prize. Relationship with Cohen-Tannoudji was built by Aspect during his doctoral research leading to a collaboration that predated the dissertation's conclusion and continued afterward.

---

[9] For the proceedings of this conference, see Progress in Scientific Culture—The Interdisciplinary Journal of the Ettore Majorana Centre, 1/4, 439–460, 1976.

## 4. The following day

Aspect's experiments marked a turning point in the history of the research on the foundations of quantum mechanics, at least as far as the issue of hidden variables and quantum mechanics is concerned. On the one hand, Aspect's experiments brought stronger evidence than previous experiments favoring quantum mechanics and opposing local hidden variables theories. On the other hand, and more importantly, these experiments brought recognition to research on these issues. Evidence of this appeared very quickly.

In 1982, just after the announcement and before the dissertation's approval, Aspect's reputation skyrocketed. He was one of the invited speakers at the Eighth International Conference on Atomic Physics, held in Sweden, to report on his experiments on Bell's inequalities. The American physicist Arthur Schawlow, Physics Nobel Prize winner in 1981, was requested to make the final report of the conference. He chose Bell's theorem and its experiments as the main topic of his speech (Schawlow 1983):

> Physical metaphors, such as the dual concepts of particles and waves in dealing with the light and atoms, are more than just conveniences, but rather are practical necessities. [. . .] But the experiments on Bell's inequalities are making it difficult for us to continue using some of our familiar physical metaphors in the old ways. We are used to thinking that light waves are produced at an atom with definite polarizations and are subsequently detected by remote detectors. However, the experiments show that if anything is propagated, it seems to convey more polarization information than a transverse wave. [. . .] As an experimentalist, I like to think that there is something there that we call an atom, and that we can make good measurements on it if we are careful not to disturb it too much. But the experiments on polarization of correlated photons don't bear out these expectations.

Two years later, Feynman, who once refused to talk about hidden variables with John Clauser while the first experiment on Bell's theorem by Freedman and Clauser was being carried out (Freire 2015, p.272), attended a seminar given by Aspect at Caltech on the tests of Bell's theorem. At this seminar,

Aspect finished his talk by quoting a certain paper whose author derived results like Bell's inequalities and went on to discuss whether it was "a real problem." According to Aspect, this author had provided an answer that was so unclear that he "had found it amusing to quote it as a kind of joke to conclude this presentation." Only at this point did Aspect reveal the name of the author, Richard Feynman. According to Aspect, nobody in the audience laughed, until Feynman himself laughed. Feynman later checked the quotation and wrote to Aspect, conceding he was right and saying, "once again let me say, your talk was excellent." [10]

Schawlow's and Feynman's positive reactions may be framed in the context of the number of citations that Aspect's papers would obtain for years to come. By any standard, these papers have had an expressive number of citations. So far, these three papers have received 4,895 citations. If we consider only a few years following the publication of these papers, the figures are rather important as they may give us a sense of the immediate recognition of Aspect's work. In the following table we have the number of citations for each paper in the years 1982, 1983, 1984, and 1985, as well as the total number of citations so far. Considering such that such papers began to get citations from roughly 1983 on, we may calculate the average number of citations, per year and per paper, from 1983 to 1985. Thus, we have the impressive figure of 27 citations per paper per year, which is an impressive number even by the current standards, 40 years later, with an enlarged number of researchers.

| Paper/citations | 1982 | 1983 | 1984 | 1985 | 1982/2022 |
|---|---|---|---|---|---|
| 1st paper | 10 | 16 | 24 | 21 | 1,185 |
| 2nd paper | 3 | 18 | 25 | 33 | 1,525 |
| 3rd paper | -- | 18 | 34 | 41 | 2,185 |

Source: Web of science, accessed on May 10, 2022. 1st paper is (Aspect et al. 1981), 2nd paper is (Aspect et al. 1982b), and the 3rd paper is (Aspect et al. 1982a)

---

[10] Richard Feynman to Alain Aspect, 28 Sep 1984. Richard P. Feynman Papers, Box 22, Folder 15, California Institute of Technology Archives. On the episode, see Freire (2015, p. 278).

In a study on the history of the research on the foundations of quantum mechanics, between 1950 and 1990, I coined the physicists who approached such a subject as quantum dissidents (Freire Junior, 2015), their dissent being related to the fact that they supported that there was good physics to be done concerning the foundations of quantum mechanics. This meant they challenged the view, shared by most of the physicists at the time, that foundational issues had already been solved by the founding fathers of the discipline. The quantum dissidents included physicists such as Bohm, Everett, Bell, Eugene Wigner, Shimony, Clauser, d'Espagnat, Aspect, H-D Zeh, Franco Selleri, and Tony Leggett. In this ensemble, however, Aspect played a singular role as a transitional protagonist. He was aware of the prevailing prejudices against the kind of research he intended to conduct. John Bell might have been the first to warn him of this when Aspect looked for him to obtain for his planned experiments. Bell asked him: "Do you have a permanent job?" Luckily Aspect did, and this made a huge difference. Initially, he suffered the indifference toward his subject but overcame it. His pedagogical skills to explain to his colleagues why testing Bell's theorem in this context was relevant for physics development helped him. However, the major transition came after the publication of the three papers. As we have seen, he was immediately acknowledged by the physicists as somebody who had done first-class physics. Some of the dissidents, however, paid a high professional price for their dissidence.[11]

**Epilogue**

As we have seen, Aspect was not optimistic about the possibility of improving his experimental technique to replicate experiments related to Bell's theorem. Indeed, his experiments had been so convincing that in subsequent years nobody else bothered to replicate them. The reasons for this were related to the perceived unfeasibility of new experimental breakthroughs, as Aspect himself remarked: "I do not see further meaningful progress can be made in the domain of Bell's inequalities, at least with our apparatus. We have exploited its maximal possibilities. Sure, an additional decimal could be obtained, but would it be worth it?"[12]. Aspect himself moved to other rewarding topics of research. Invited by Cohen-Tannoudji, he began to work on the use of lasers to cool down atoms and later became a leader in the field of atomic optics.

---

[11] See, particularly, the cases of Everett, Clauser, Tausk, and Zeh, in Freire (2015).

However, the story about experimental tests of Bell's theorem did not end with Aspect's work. The revival began slowly five years later with the discovery of a better way to produce pairs of photons with entangled polarizations. The new sources did not use atomic cascades, instead, the pair of entangled photons were created in the interaction between a laser beam and nonlinear optical crystals, a process named parametric down-conversion (PDC). While this process was known since the early days of nonlinear optics, only in the late 1980s, Yanhua Shih and Carroll Alley at the University of Maryland in College Park and Zhe-Yu Ou and Leonard Mandel at the University of Rochester (Shih & Alley 1988, Ou & Mandel 1988) had the idea to use this source to redo experiments with Bell's theorem. The first experiment with the new source did not produce better violations of Bell's inequalities than those previously obtained by Alain Aspect. However, as time has gone by, the use of this source was improved, both in conceptual as well experimental aspects, and the results surpassed Aspect's results. In addition to this better source of pairs of entangled photons, there remained loopholes other than the locality, which had been addressed by Aspect. Physicists were aware, since the first experiments on Bell's theorem, that such loopholes allowed the survival of the local realism assumption even if they were hardly plausible.

The loopholes were related to some additional assumptions experimental physicists need to take to make the transition from Bell's original inequality to the CHSH's one, which could be applied to a factual experiment, and to real lab experiments. The locality loophole, as we have seen, was approached by Aspect's third experiment, but the quasi-random nature of the changing polarizers was a limitation to consider it had been closed. Another loophole, called 'detector-efficiency loophole' or 'fair-sampling loophole', derived from the fact that as we were not able to detect all the pairs of photons, the sample taken for the statistical calculations to compare experimental results with Bell's inequalities, could be, in principle, at least, biased. Indeed, with detectors used in the early

---

[12] Aspect, interviewed in Deligeorges (1985, p. 137)

Bell's theorem experiments, it was possible to mimic experimental results departing from local realist assumptions. The third loophole has been called the freedom-of-choice. While the locality loophole concerns the transmission of information among parts of the experimental device, the freedom-of-choice loophole means the possibility of a hypothetical common cause interfering in statistical correlation among the entangled systems. These loopholes have been tackled by different teams, with different strategies, and in different places.[13]

A review of the wide series of Bell's theorem experiments dealing with these loopholes falls beyond the scope of this paper. A few milestones in this series were the following: In 1998, Anton Zeilinger and his team improved on Aspect's 1982 experiment with time-varying analyzers reinforcing the condition of locality by changing the analyzers in a stronger random manner with the detectors separated by 400 m. They got a full agreement with quantum mechanics predictions and violations of Bell's inequality by over 30 standard deviations (Weihs et al. 1998). In 2015, three different experiments closed, at the same time, the locality, and the fair-sampling loopholes. They were held in Austria, the US, and the Netherlands, led by Zeilinger, at the University of Vienna, Lynden Shalm at NIST in Boulder, Colorado, and Ronald Hanson, at Delft University of Technology. Zeilinger's and Shalm's teams took advantage from the newly available photon detectors with efficiency above 90%, in addition to using pairs of entangled photons through parametric down conversion and a scheme with a new type of random number generator to change the polarizer's alignments. Hanson's team developed a different strategy to prevent the fair-sampling loophole. Their scheme, called "Bell's event-ready scheme", allowed them to measure spin components from a kind of artificial atom consisting of two nitrogen-vacancy (NV) centers. This scheme was further developed by H. Weinfurter and colleagues in Garching, Germany. The impact of the 2015 three

---

[13] For a conceptual presentation of these three loopholes, as well as for a survey of recent Bell's experiments, see Kaiser (2022). On the history of techniques related to Bell's theorem experiments, see (Silva Neto, 2022)

experiments was assessed by Aspect (2015) in a "Viewpoint" paper meaningfully titled "Closing the Door on Einstein and Bohr's Quantum Debate."[14]

Aspect's title was right if we consider the third loophole was not on the agenda of the debate between Einstein and Bohr. However, as a logical possibility, initially noted by Shimony, Horne, and Clauser, in 1976; it required to be addressed.[15] Jason Gallicchio, Andrew Friedman, and David Kaiser (2014) suggested to use signals from cosmic sources to change the polarizer's alignments and, through this strategy, to send backward the time of that hypothetical common cause. This proposal led to a joint effort among physicists in Austria and in the US, led by Zeilinger. Eventually, they were able to use signals coming from two different quasars to adjust their setting, which took 7,78 billion years to arrive in one of the detectors 12,21 to the other. Thus, as concluded by Kaiser (2022, 361), this experiment "excluded such local-realist, freedom-of-choice scenarios from 96:0% of the space-time volume of the past light cone of the experiment, extending from the Big Bang to the present time," thus corroborating the current view that quantum entanglement is a true physical nature feature.[16]

Parallel to the new experiments with entanglement, there was yet another surprise in store for Aspect. Since late 1980s, physicists began to conjecture about the use of entanglement, the physical effect brought to light by research on Bell's theorem for quantum computing and cryptography. At a certain juncture, in the mid-1990s, new fields in physics were born: quantum information and cryptography. The core physical effects in these fields are entanglement and decoherence. Thus, the 2010 Wolf Prize in physics, awarded to Clauser, Aspect, and Zeilinger, recognized the role played by the leaders of the three distinct generations of physicists who had worked on Bell's theorem and paved the road to the current blossoming of quantum information.

---

[14] Vienna's experiment is (Giustina et al, 2015), NIST's experiment is (Shalm et al, 2015), Delft's experiment is (Hensen et al, 2015), and Garching's experiment is (Rosenfeld et al, 2017).

[15] For the history of the freedom-of-choice loophole, and attempts to close it, see Kaiser (2022, from p. 356 on).

[16] The cosmic experiment's results are in (Rauch et al., 2018). See also (Kaiser, 2022, p. 361) for the experiment conducted in Shanghai by Jian-Wei Pan and his team (Li et al., 2018), where they closed the three loopholes in the same investigation while taking signals from nearby bright stars.

**Data Availability Statement**

The datasets generated during and/or analyzed during the current study, such as letters, interviews, and citations, are available from the corresponding author on reasonable request.